\documentclass[useAMS,usenatbib]{mn2e}
\usepackage{graphicx}
\usepackage{longtable}

\title[On a possible Ceres paleo-family]
{Footprints of a possible Ceres asteroid paleo-family}
\author[V. Carruba, D. Nesvorn\'{y}, S. Marchi, S. Aljbaae]{V. Carruba$^{1,2}$\thanks{E-mail: vcarruba@feg.unesp.br}, D. Nesvorn\'{y},$^{2}$, S. Marchi$^{2}$, 
S. Aljbaae$^{1}$\\
$^{1}$UNESP, Univ. Estadual Paulista, Grupo de din\^{a}mica Orbital e
  Planetologia, Guaratinguet\'{a}, SP, 12516-410, Brazil. \\
$^{2}$Department of Space Studies, Southwest Research Institute, Boulder, 
CO, 80302, USA.\\
}

\begin{document}

\date{Accepted 2016 February 15.  Received 2016 February 15; in original form 2016 January 8.}

\pagerange{\pageref{firstpage}--\pageref{lastpage}} \pubyear{2016}

\maketitle

\label{firstpage}

\begin{abstract}
Ceres is the largest and most massive body in the asteroid main belt.  
Observational data from the Dawn spacecraft reveal the presence of at least
two impact craters about 280~km in diameter on the Ceres surface, 
that could have expelled a significant number of fragments.  
Yet, standard techniques for identifying dynamical asteroid families have 
not detected any Ceres family.  In this work, we argue that linear 
secular resonances with Ceres deplete the population of objects near Ceres. 
Also, because of the high 
escape velocity from Ceres, family members are expected to be very 
dispersed, with a considerable fraction of km-sized fragments that should 
be able to reach the pristine region of the main belt, the area between the 
5J:-2A and 7J:-3A mean-motion resonances, where the observed number 
of asteroids is low.  Rather than looking for possible Ceres family members 
near Ceres, here we propose to search in the pristine region.  We identified 
156 asteroids whose taxonomy, colors, albedo  could be compatible with being 
fragments from Ceres.  Remarkably, most of these objects have 
inclinations near that of Ceres itself.  

\end{abstract}

%

\section{Introduction}
\label{sec: intro}
Ceres is the largest and most massive body in the asteroid main belt.  
While many other large bodies of similar composition such as Pallas, 
Hygiea, Euphrosyne have asteroid families, no such group has been so 
far identified for Ceres \citep{Milani_2014, Nesvorny_2015}.  Yet collisional 
models suggest that about 10 craters larger than 10 km in diameter
should have formed over 4.55 Gyr of collisional evolution in the main 
belt \citep{Marchi_2016}.  
Also, observational data from the Dawn probe show that at least two 
$\simeq$280-km sized craters were formed in the last $\simeq 2$ Gyr on 
Ceres surface, and larger impacts may have happened in the past 
\citep{Marchi_2016}.  The absence of a Ceres family is therefore a 
major mystery in asteroid dynamics \citep{Milani_2014, Rivkin_2014, 
Nesvorny_2015}. Previous works have suggested the possibility that the outer
shell of Ceres could have been rich in ice, and that family members could 
have sublimated or been eroded by collisional evolution on timescales of 
hundreds of millions of years \citep{Rivkin_2014}.  Alternatively,
it has also been hypothesized that km-sized fragments ejected at Ceres
escape velocity would likely disintegrate \citep{Milani_2014}.
Results from the Dawn mission set however an upper limit of 40\% in the 
ice content of Ceres outer shell \citep{DeSanctis_2015, Fu_2015},
which seems too low to explain the lack of a family.

In this work we argue that the unique nature of Ceres as a dwarf planet may 
have indeed produced families not easily detectable using methods 
calibrated for smaller, less massive bodies. Close encounters and 
linear secular resonances with Ceres are expected to have significantly 
depleted the orbital region in the near proximity of this body, so reducing 
the number of the closest Ceres neighbors.  More importantly, initial 
ejection velocities should have been significantly larger than those 
observed for any other parent body in the main belt, including Vesta, 
spreading the collision fragments in a much larger area.  Members of the 
Ceres family would be significantly more distant among themselves than the 
typical distances between objects formed in collisions from smaller bodies, 
making an identification of the Ceres family quite challenging.

Rather than looking for members of a Ceres group in the central main belt, 
the region of the main belt with the highest concentration of asteroids, 
and where two other large Ch families, Dora and Chloris, 
\citep{Nesvorny_2015}, would make the taxonomical identification of former 
members of a Ceres family or paleo-family (an old dispersed family with an age 
between 2.7 and 3.8 Gyr, \citet{Carruba_2015b}) difficult, in this work we 
suggest to search for Ch-type objects in the pristine region of the main belt, 
between the 5J:-2A and 7J:-3A mean-motion resonances (or between 2.825 and 
2.960 au in proper semi-major axis).  This region of the 
asteroid belt was depleted of asteroids during the Late Heavy Bombardment
(LHB hereafter) phase by sweeping mean-motion resonances, and the two 
mean-motion resonances with Jupiter have since limited the influx of 
outside material from other areas of the asteroid belt \citep{Broz_2013}.  The 
lower density of asteroids and the lack of other major C-type families at 
eccentricities and inclinations comparable to those of Ceres makes the 
identification of possible members of the Ceres family an easier task in 
this region.  Identifying concentrations of C-type objects at values of 
inclinations comparable with that of Ceres itself could therefore be a strong
circumstantial evidence about the formation of a Ceres family in the
past.

\section{Effects of local dynamics} 
\label{sec: Ceres_dyn}

\begin{figure*}

  \centering
  \begin{minipage}[c]{0.45\textwidth}
    \centering \includegraphics[width=3.in]{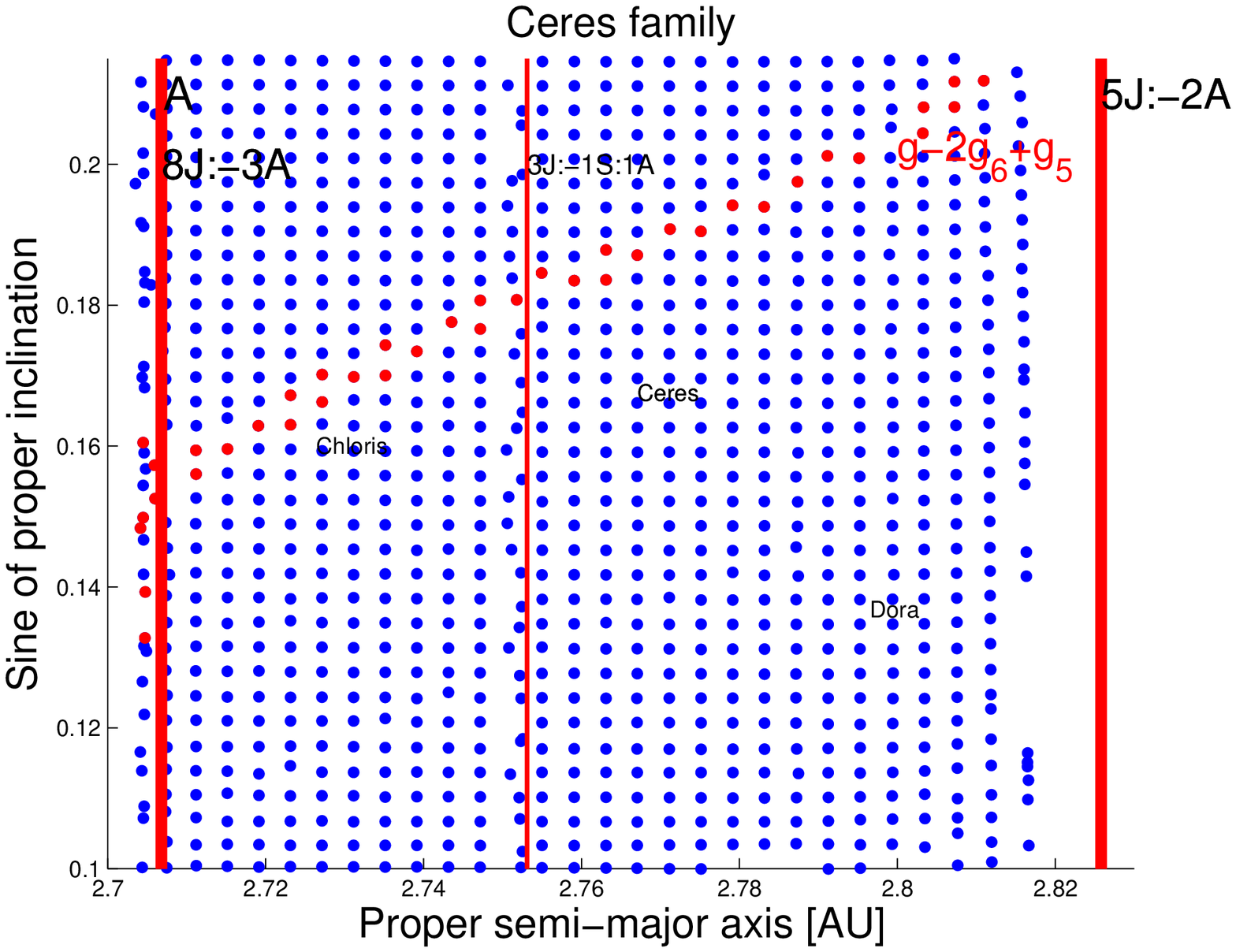}
  \end{minipage}%
  \begin{minipage}[c]{0.45\textwidth}
    \centering \includegraphics[width=3.in]{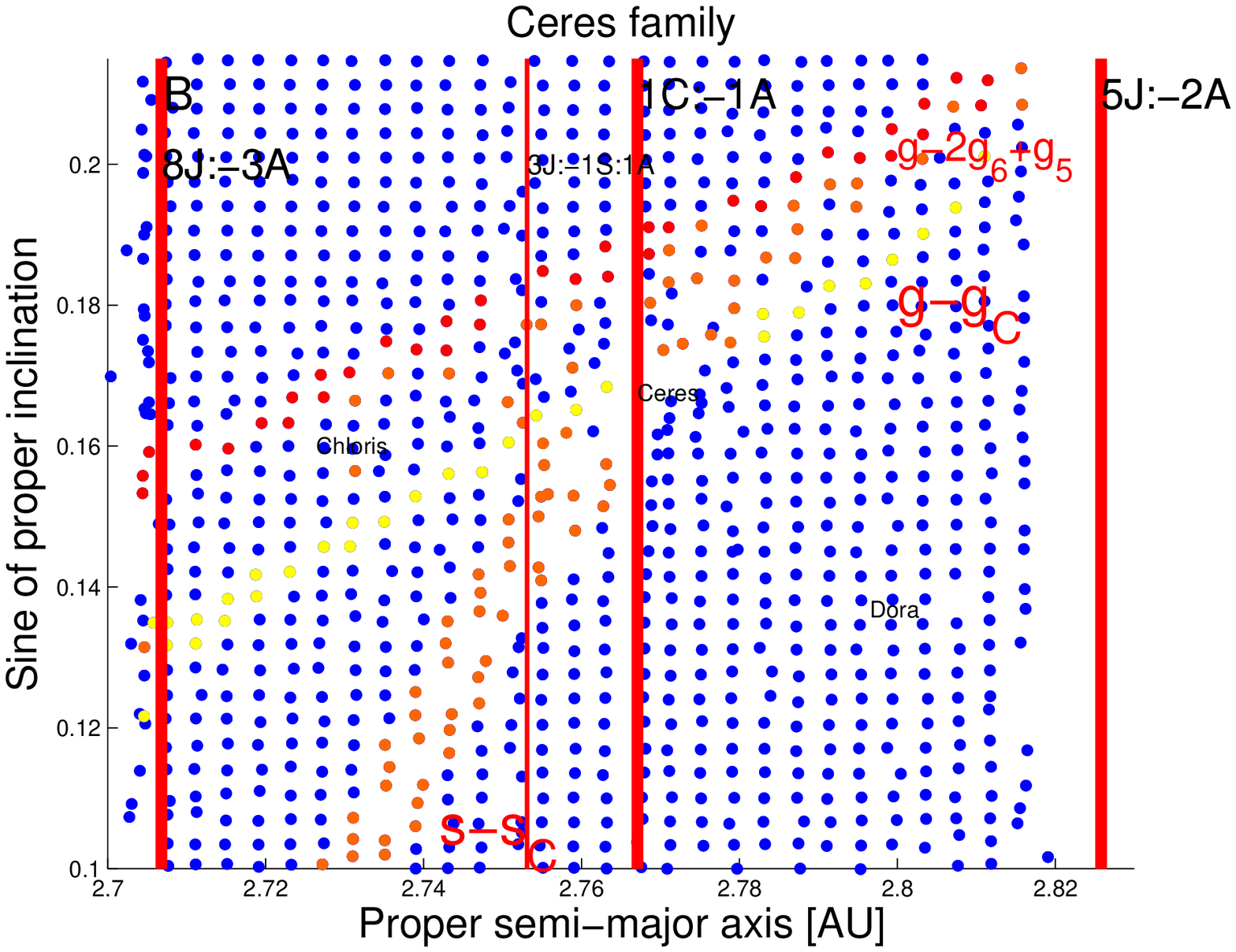}
  \end{minipage}

\caption{Dynamical maps for the orbital region of Ceres obtained by
integrating test particles under the influence of all planets (panel A), 
and all planets and Ceres as a massive body (panel B). Unstable
regions associated with mean-motion resonances appear as vertical  
strips.  Secular resonance appear as inclined bands of 
aligned dots.  Dynamically stable regions are shown as uniformly 
covered by blue dots.  Vertical lines display the location of the main 
mean-motion resonances in the area.  Red filled dots in both panels show 
the locations of ``likely resonators'' in the $g-2g_6+g_5$ secular 
resonance.  Yellow and orange filled dots in panel B show the orbital 
locations of likely resonators in the linear secular resonances with 
Ceres of pericenter and node, respectively.  The orbital location of 1 
Ceres, 668 Dora, and 410 Chloris are labeled.} 
\label{fig: map_central}
\end{figure*}

Ceres is the only dwarf planet in the asteroid belt. While its
mass is not large enough to clear its orbital neighborhood, 
significant dynamical perturbations are expected from this body, either 
as a consequence of close encounters \citep{Carruba_2003} or because of linear
secular resonances involving the precession frequency 
of Ceres node $s_C$ or pericenter $g_C$, and those of a given 
asteroid \citep{Novakovic_2015}.  To investigate the importance
of Ceres as a perturber, we obtain proper elements for
1400 particles in the plane of proper semi-major axis and inclination 
$(a,\sin{(i)})$, with the approach described in \citet{Carruba_2010},
based on the method for obtaining synthetic proper elements of
\citet{Knezevic_2003}.  We integrated the particles over 20 Myr under 
the gravitation influence of i) all planets and ii) all planets plus Ceres as a
massive body\footnote{The mass of Ceres was assumed to be equal
to $9.39 \cdot 10^{20}$~kg, as determined by the Dawn spacecraft 
\citep{Russell_2015}.} with $SWIFT\_MVSF$, the symplectic integrator
based on $SWIFT\_MVS$ from the {\em Swift} package of \citet{Levison_1994},
and modified by \citet{Broz_1999} to include online filtering of osculating
elements.  The initial osculating elements of the particles went
from 2.696 to 2.832~au in $a$ and from $5.5^{\circ}$ to $13.3^{\circ}$ in 
$i$.  We used 35 intervals in $a$ and 40 in $i$.  The other 
orbital elements of the test particles were set equal to 
those of Ceres at the modified Julian date of 57200.

Fig.~\ref{fig: map_central} displays our results, as two
dynamical maps.  Each blue dot is associated with a particle proper 
$(a,\sin{(i)})$ value.  For the
case without Ceres as a massive body, displayed in panel A, the orbital region
near Ceres is relatively stable. The only important non-linear secular resonance
in the region is the $g-2g_6+g_5$ (or $2{\nu}_6-{\nu}_5$, in terms of 
linear secular resonances).  Objects whose pericenter frequency is within
$\pm0.3$ arcsec/yr from $2g_6+g_5 = 52.229$ arcsec/yr, likely resonators
in the terminology of \citet{Carruba_2009}, are shown as red full dots in this
figure.  This is a diffusive secular resonance, and asteroids captured
into this resonance may drift away, but will not be destabilize 
\citep{Carruba_2014}.  The situation is much different and more 
interesting for the map with Ceres as a massive body.

As observed in Fig.~\ref{fig: map_central}, panel B, accounting for
Ceres causes the appearance of a 1:1 mean-motion resonance
with this dwarf planet.  Most importantly, linear secular resonances of nodes
$s-s_C$ and pericenter $g-g_C$, first detected by \citet{Novakovic_2015}
significantly destabilize the orbits in the proximity of Ceres.  Combined
with the long-term effect of close encounters with Ceres, this has 
interesting consequences for the survival of members of the Ceres family in
the central main belt.  Not many family members are expected to survive
near Ceres, and this would cause significant difficulties in 
using standard dynamical family identification techniques, since they are based
on looking for pairs of neighbors in proper element domains 
(Hierarchical Clustering Method, HCM hereafter, \citet{Bendjoya_2002}).  
Since the close neighbors of Ceres would have 
been removed on a short timescale, only objects
whose distance from Ceres is higher than the average distance between pairs of 
asteroids in the central main belt would have survived.   
Also, the area at higher inclination than that associated with the 
linear secular resonances with Ceres should be significantly depleted of 
asteroids, and this is actually observed (see for instance Fig.~9 in 
\citet{Nesvorny_2015}).  Finally, the Dora family, a large ($\simeq 1260$ 
members, \citet{Nesvorny_2015}) group of the same spectral type of Ceres, Ch, 
is located at lower inclinations than that of Ceres.
Considering that the large Ch Chloris family is also encountered in the
central main belt near Ceres, discriminating between a Ch asteroid
from Ceres and one from Dora, Chloris, or other local sources would be a 
very daunting task, also considering 
that there are currently no identified members of a Ceres family that 
could serve as a basis for a taxonomical analysis.

To further investigate the efficiency of Ceres in perturbing
its most immediate neighbors, we also performed numerical simulations
of a fictitious Ceres family with 
the $SYSYCE$ integrator (Swift$+$Yarkovsky$+$Stochastic
YORP$+$Close encounters) of \citet{Carruba_2015}, modified to also
account for past changes in the values of the solar luminosity.
This integrators accounts for both the diurnal and seasonal
version of the Yarkovsky effect, the stochastic version of the YORP effect,
and close encounters of massless particles with massive bodies.
The numerical set-up of our simulations was similar to what discussed
in \citet{Carruba_2015}: we used the optimal values of the Yarkovsky 
parameters discussed in \citet{Broz_2013} for C-type asteroids,
the initial spin obliquity was random, and normal reorientation timescales 
due to possible collisions as described in \citet{Broz_1999} were 
considered for all runs. We integrated our test particles  
under the influence of all planets (case A), and all planets plus Ceres
(case B), and obtained synthetic proper elements
with the approach described in \citet{Carruba_2010}.

\begin{figure*}

  \centering
  \begin{minipage}[c]{0.45\textwidth}
    \centering \includegraphics[width=3.in]{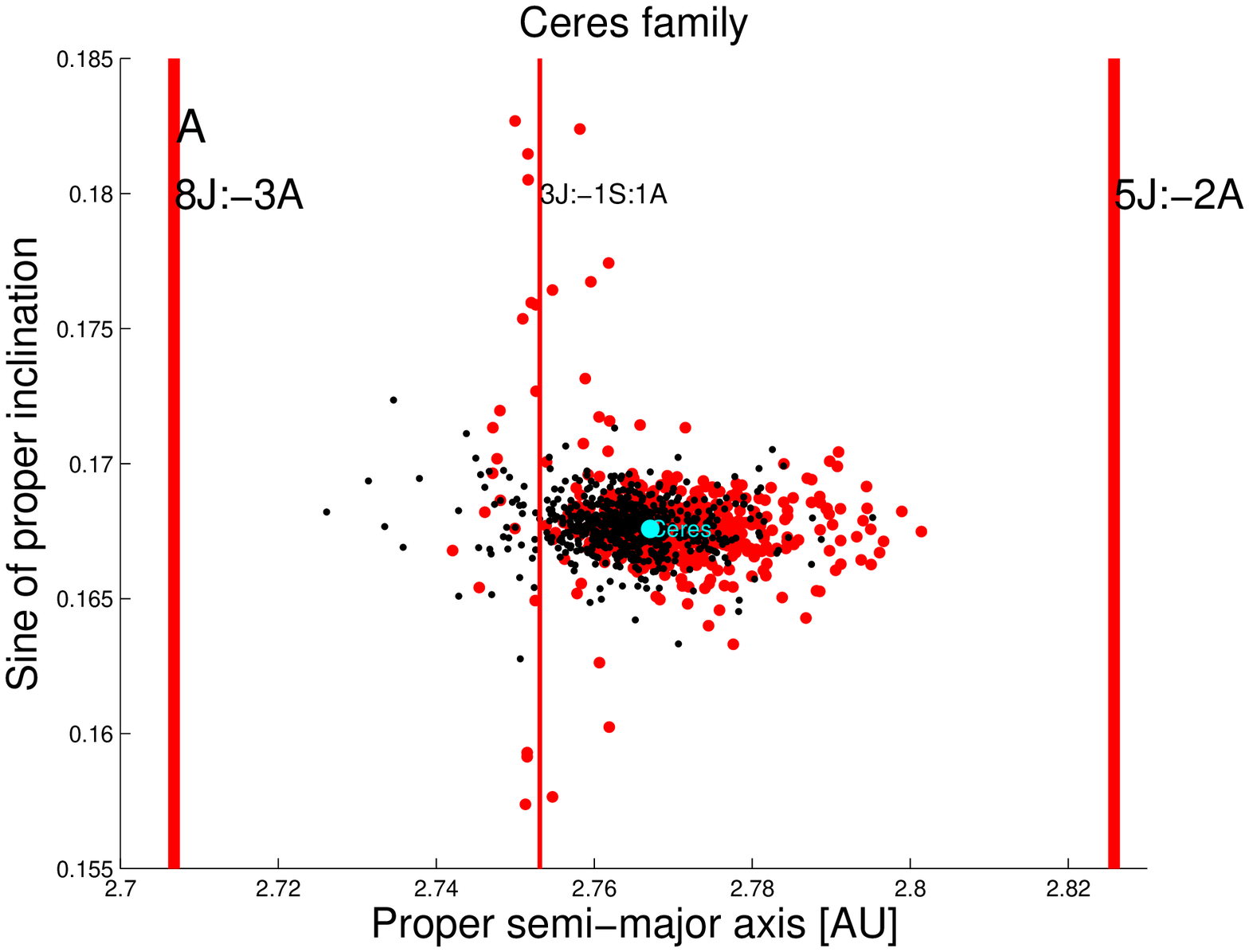}
  \end{minipage}%
  \begin{minipage}[c]{0.45\textwidth}
    \centering \includegraphics[width=3.in]{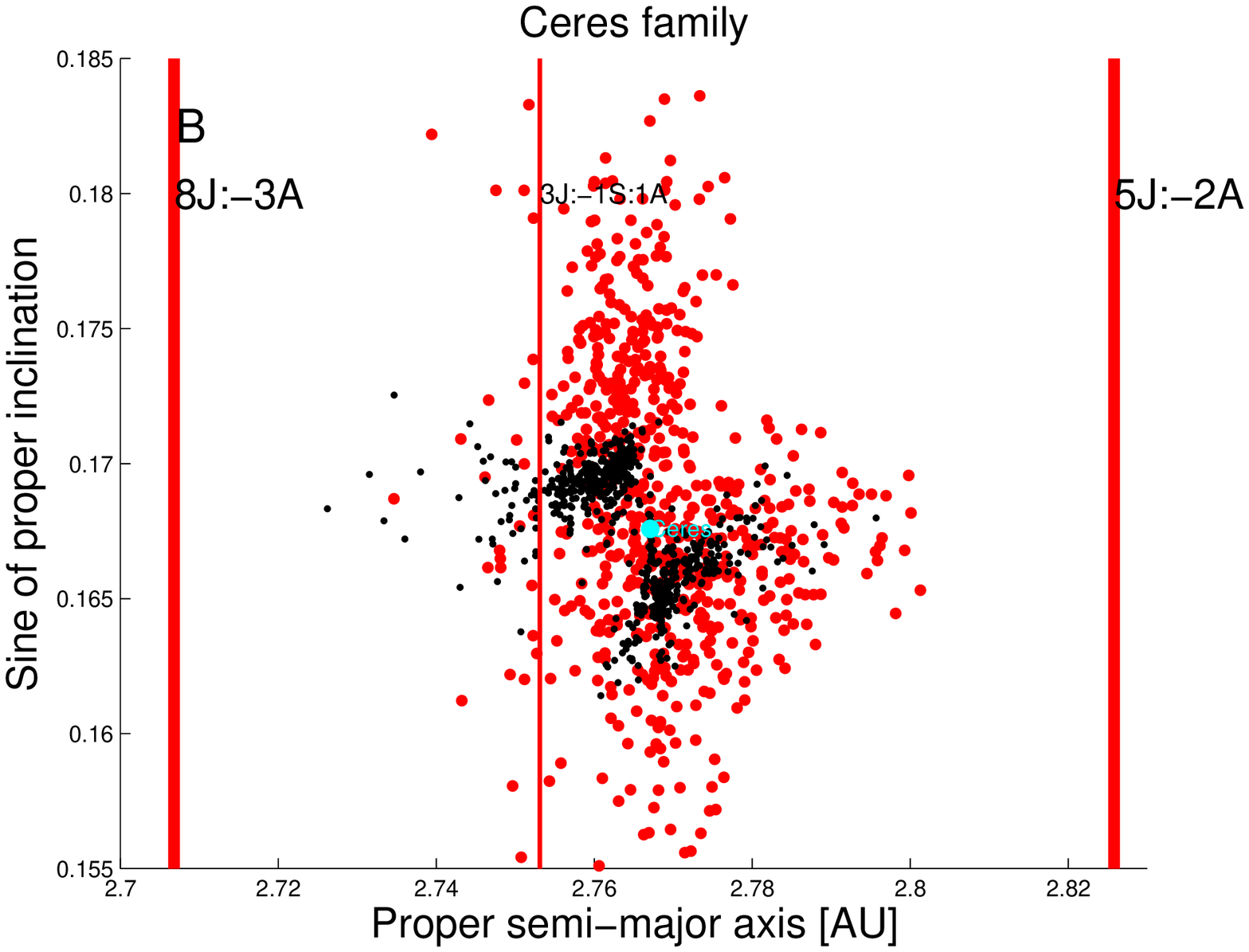}
  \end{minipage}

\caption{Projection in the proper $(a,\sin{(i)})$ plane of the initial (black
full dots, $T = 1.2$~My) and final (red full dots, $T= 455.6$~My) elements
of the simulated Ceres family members integrated under the gravitational
influence of all planets (panel A), and all planets plus Ceres (panel
B).  Vertical lines display the location of the mean-motion resonances,
the cyan full dot identify the orbital location of Ceres itself.} 
\label{fig: ceres_scatter}
\end{figure*}

The fictitious Ceres family was generated with the approach described in 
\citet{Vokrouhlicky_2006}: we assumed that the initial ejection velocity
field follows an isotropical Gaussian distribution of zero mean and 
standard deviation given by:

\begin{equation}
{\sigma}_{V_{ej}}=V_{EJ}\cdot \frac{5km}{D},
\label{eq: std_veJ}
\end{equation}

\noindent 
where $D$ is the body diameter in km, and $V_{EJ}$ is a parameter 
describing the width of the velocity field.  We used the relatively small
value of $V_{EJ} = 60$~m/s for our simulated family.  In the next section
we will discuss why it is quite likely that any possible Ceres family
would have a much larger value of $V_{EJ}$.  Since here, however,
we were interested in the dynamics in the proximity of Ceres itself,
our choice of $V_{EJ}$ was, in our opinion, justified.   We generated 
642 particles with size-frequency distributions (SFD) with an 
exponent $-\alpha$ that best-fitted the cumulative distribution equal to 
3.6, a fairly typical value \citep{Masiero_2012}, and with diameters in 
the range from  2.0 to 12.0 km.

Fig.~\ref{fig: ceres_scatter} displays our results in the proper 
$(a,\sin{(i)})$ plane for the simulation without Ceres as a massive 
body (panel A) and with a massive Ceres (panel B).  Black full dots 
display the position of simulated Ceres family members at the beginning of 
the simulation, after 1.2 My, while the red full dots show the orbital 
location at the end of the integration, for $T = 455.6$~My. For the case
without Ceres one can notice that the family does not disperse much,
and that most of the scattering is caused by the 3J:-1S:1A three-body 
resonance.  Results including Ceres as a perturber are quite more 
interesting: already after just 1.2 My secular resonances with Ceres cleared
a gap near the orbital location of the dwarf planet, identified by a full
blue dot.  The situation is even more dramatic at the end of the simulation,
where we observe that Ceres scattered material at lower and higher 
inclinations with respect to the case in which Ceres had no mass, and opened
a significant gap at the family center, as also observed in the dynamical map
of Fig.~\ref{fig: map_central}, panel B.

\begin{figure}
  \centering
  \centering \includegraphics [width=0.45\textwidth]{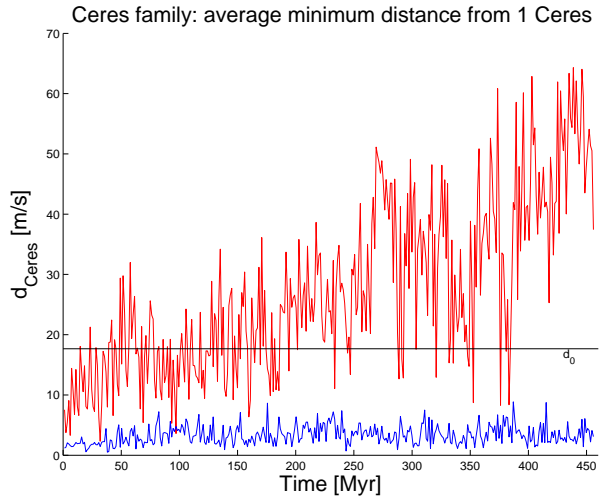}

\caption{Averaged minimum distance from Ceres $d_{Ceres}$ as a function 
of time of the simulated family members for the case without (blue line) 
and with (red line) Ceres as a massive body.  The horizontal black line displays
the maximum nominal distance cutoff level $d_0$ for the whole simulated Ceres 
family.}
\label{fig: dist_ceres}
\end{figure}

How difficult would be to identify our simulated Ceres family using HCM?
To answer this question, we computed the nominal distance velocity cutoff 
$d_0$ for which two nearby asteroids are considered to be related using 
the approach of \citet{Beauge_2001}, that defines this quantity as the 
average minimum distance between all possible asteroid pairs, as a 
function of time, for i) all paired asteroids in the simulated
Ceres family ($d_0$), and ii) just with respect to Ceres itself ($d_{Ceres}$).
This latter quantity provides an estimate of the minimum value
of distance cutoff needed to identify a Ceres family using the standard
HCM.  The maximum value of the nominal distance velocity cutoff for the
simulated Ceres family was equal to 17.7~m/s. When Ceres was not considered
as a massive body, values of $d_{Ceres}$ were of the order of 5~m/s, at all
times below the nominal distance velocity cutoff (see 
Fig.~\ref{fig: dist_ceres}, blue line).  Once Ceres was included as a 
massive body, however, values of $d_{Ceres}$ quickly surpassed the 
nominal distance velocity cutoff $d_0$ (see red line in 
Fig.~\ref{fig: dist_ceres}): after 200 My, except of short episodes in
which $d_{Ceres} < d_0$, for most of the time $d_{Ceres}$ was larger than 
$d_0$.  After 380 My, this trend was consolidated and $d_{Ceres}$ was 
always larger than  $d_0$.  The identification of a Ceres family using
HCM in the domain of proper elements of the simulated family and Ceres 
as the first body would be quite challenging after 380 My.

So, where to look for hypothetical surviving
members of a Ceres family?  The next section will be dealing with estimating
the extent of the initial orbital dispersion of a putative family, and in 
assessing the regions where locating its members could be easier.

\section{Ejection velocities of a possible Ceres family} 
\label{sec: Ceres_ej}

The escape velocity $V_{esc}$ from Ceres is considerably larger than that of 
other, less massive bodies:  $\simeq 480$ m/s \citep{Russell_2015} 
versus the 360 m/s for Vesta \citep{Russell_2012},
the largest body in the main belt with a uncontroversial 
recognized asteroid family.  The Vesta family is 
one of the largest in the main belt, and one would expect a 
putative Ceres family to be even larger.

Recently, \citet{Carruba_2016} investigated the shape of the ejection 
velocity field of 49 asteroid families.  Typical observed values
of $\beta = V_{EJ}/V_{esc}$, with $V_{EJ}$ given by Eq.~\ref{eq: std_veJ}, 
are between 0.5 and 1.5 (see also \citet{Nesvorny_2015}, sect. 
5)\footnote{Isotropical ejection velocity
field would be expected for large basins, as observed for the Vestoids.  
Cratering events, however may produce asymmetrical initial velocity field 
\citep{Novakovic_2012, Marchi_2001}, and depending on
the orbital configuration of Ceres at the time of the family formation
event, namely, its values of true anomaly $f$ and argument of pericenter
$\omega$, and on the impact geometry, family members may have not reached the 
pristine region.  For simplicity, in this work we concentrated our analysis on 
large basins forming events.  
However, since collisional models predict that about 
10 400 km or larger craters could have formed on Ceres 
\citep{Marchi_2016}, we expect
that, at least for a fraction of these events, some of the ejected
material was likely to have reached the pristine region.}.  
While for typical asteroid families this
implies that $V_{EJ}$ is generally lower than 100 m/s, even a conservative
choice of $\beta = 0.5$ for Ceres would produce a field with $V_{EJ}$ 
of the order of 240 m/s.  This is more than a factor of 2 larger than 
the higher values of $V_{EJ}$ observed for some of the oldest asteroid
families \citep{Carruba_2015b}, but not inconsistent with the observed
presence of crater ray systems around large craters on Ceres, such 
as Occator \citep{Nathues_2015}, that cover a large portion of the Ceres 
surface.  Since the circular velocity needed to form such rays is of the 
order of half the escape velocity from Ceres, this suggests that values of 
$\beta$ near 0.5 for the $V_{EJ}$ parameter of a possible Ceres family are 
not unreasonable.  Following the approach of \citet{Vokrouhlicky_2006}, see 
also \citet{Carruba_2015b}, we generated fictitious Ceres asteroids members 
with diameters 1, 3, and 7~km  with the minimum and maximum value of $\beta$ 
from \citet{Carruba_2016}.

\begin{figure*}

  \centering
  \begin{minipage}[c]{0.45\textwidth}
    \centering \includegraphics[width=3.in]{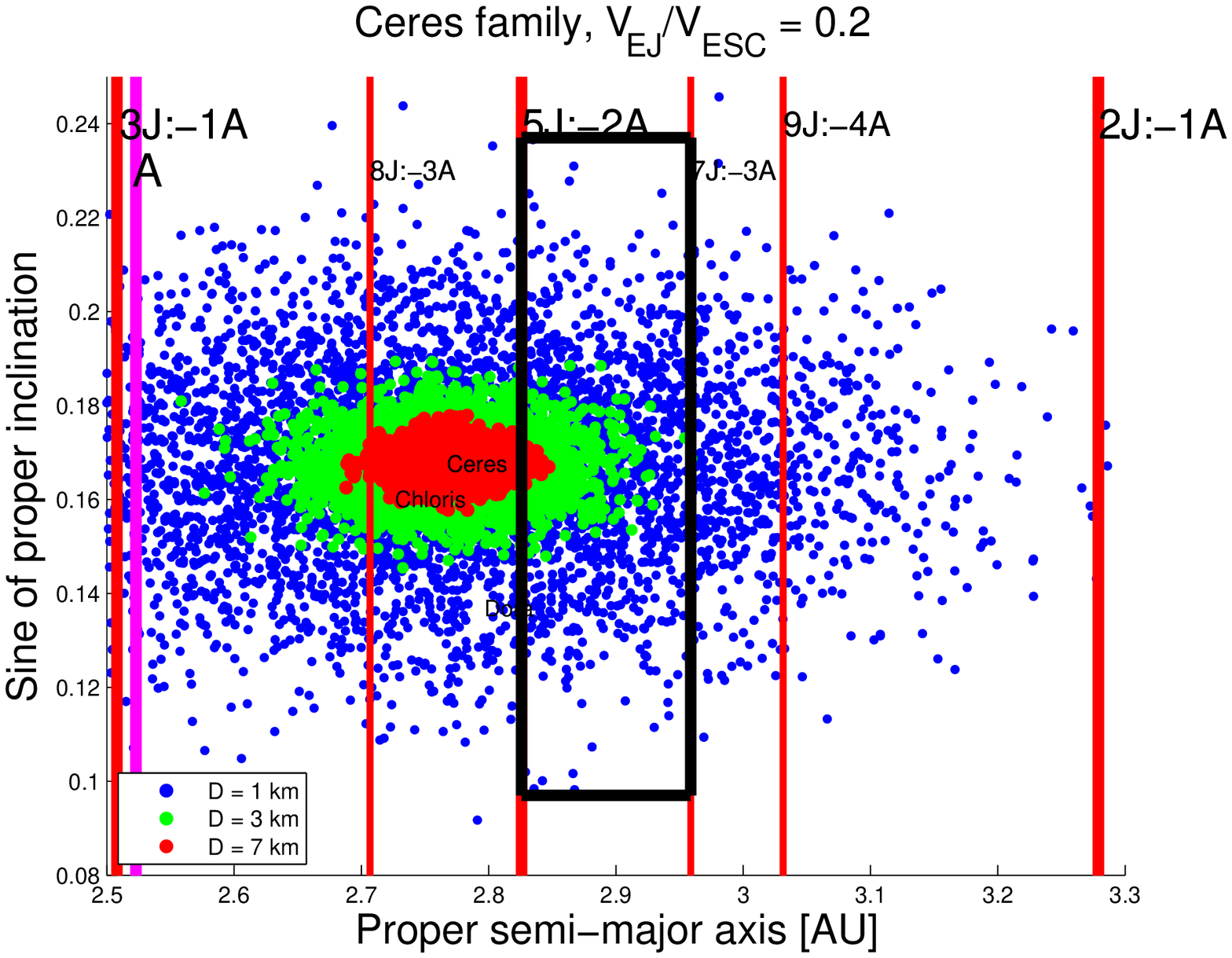}
  \end{minipage}%
  \begin{minipage}[c]{0.45\textwidth}
    \centering \includegraphics[width=3.in]{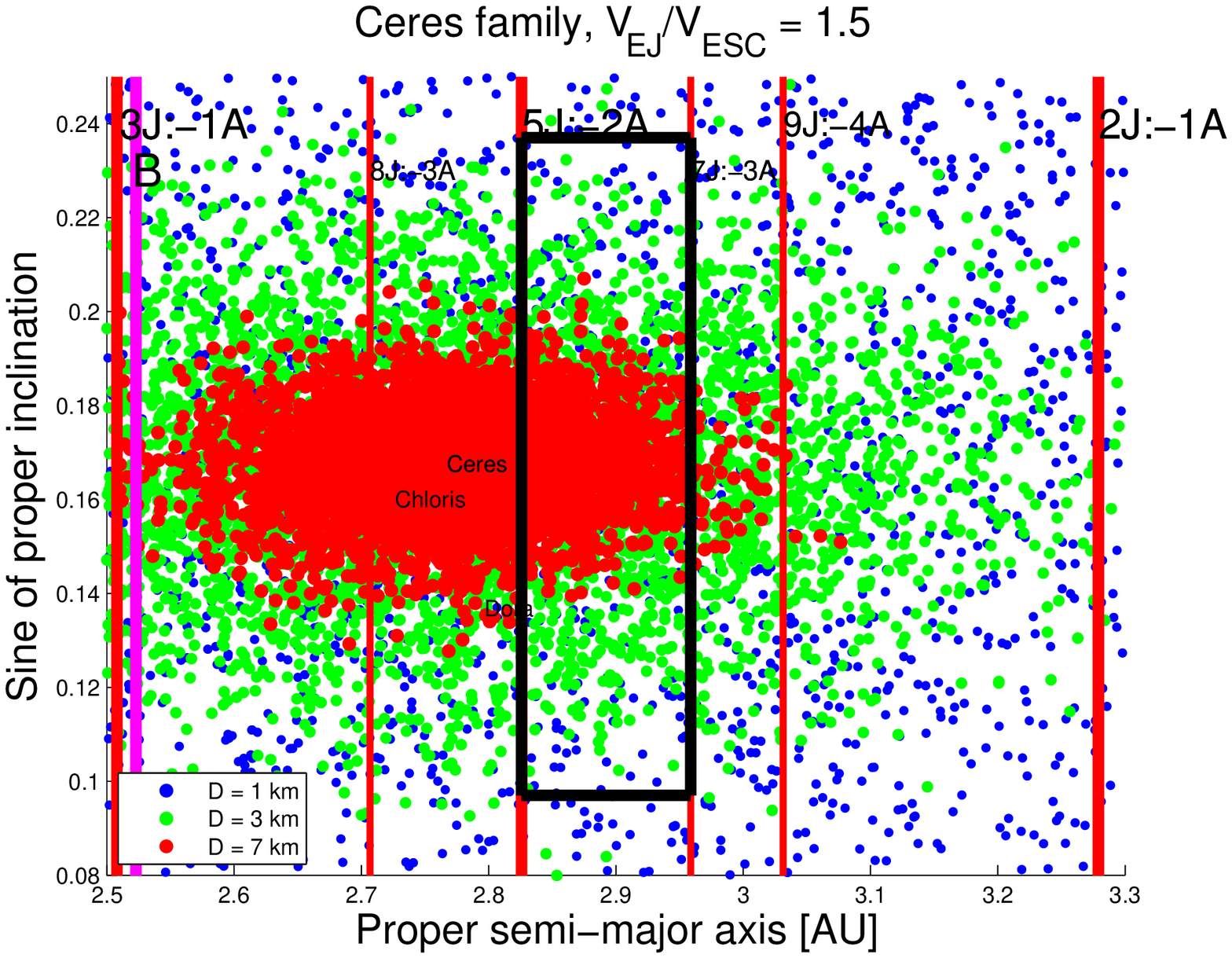}
  \end{minipage}

\caption{An $(a,sin(i))$ distribution of $D= 1, 3,$ and $7$~km members
of Ceres families obtained using $V_{EJ} = 0.2 V_{ESC}$ (panel A)
and $V_{EJ} = 1.5 V_{ESC}$ (panel B).   The black box 
identify the orbital regions in which most of the $D = 3$ km 
fragments for $V_{EJ} = 1.5 \cdot V_{ESC}$ and most of the $D = 1$ km 
for $V_{EJ} = 0.2 \cdot V_{ESC}$ could be found in the 
pristine region of the main belt.} 
\label{fig: ceres_ej}
\end{figure*}

Fig.~\ref{fig: ceres_ej} displays our results for $\beta = 0.2$ (panel 
A) and $\beta = 1.5$ (panel B).  Even in the most 
conservative case of $\beta = 0.2$, a value lower than those of any of the 
families studied in \citet{Carruba_2016}, we still found that
a significative fraction\footnote{34.2, 10.6, and 0.5\% of 
the $D = 1, 3,$ and $7$~km populations, respectively.} of the multi-km 
(and potentially observable) fragments from Ceres could have been
injected in the pristine region of the main belt.  We define this region
as the area between the 5J:-2A and 7J:-3A mean-motion resonances with 
Jupiter.  Because of these two dynamical barriers, very little material
from external regions can reach this area \citep{Broz_2013}.  As a consequence,
the local density of asteroids is much inferior to that of other regions
of the asteroid belt.  Also, only small C-type families, such those of 
Naema (301 members), Terpsichore (138 members), and Terentia (79 members)
\citep{Nesvorny_2015} are found in the region, and at eccentricity
values quite different than that of Ceres, so easily distinguishable
from putative Ceres fragments.  If any large
($D > 5$~km), not easily movable by non-gravitation effects, members of a 
hypothetical Ceres family would have been injected in this region 
after the LHB\footnote{The LHB likely occurred 4.1-4.2 Gyr ago 
(e.g., \citet{Marchi_2013}), no family is expected 
to have survived this event, and we also expect he original population
of asteroids in the pristine region to have been significantly depleted
\citep{Brasil_2015}.}, it would be reasonable to expect that they 
could have remained there and still be visible
today.  We selected a region in the pristine zone where most of the 
$D > 1$~km objects could be found, according to our results for
the most conservative case with $\beta = 0.2$ (black box
in Fig.~\ref{fig: ceres_ej}, panels A, this also
corresponds to the region in which most of the $D > 3$~km objects
would be located in the most optimistic scenario of $\beta = 1.5$, 
see black box in Fig.~\ref{fig: ceres_ej}, panels B.  In the next 
section we will investigate if any such object could be identified in 
the pristine region.

\section{The pristine region} 
\label{sec: Ceres_pristine}

Results of a dynamical map with Ceres as a massive body, not 
shown for the sake of brevity, demonstrate that the pristine region at 
inclination close to that of Ceres is dynamically stable, with 
no linear secular resonances with Ceres and only minor diffusive secular 
resonances with the giant planets.  
In the black region of Fig.~\ref{fig: ceres_ej} there are four Ch asteroids, 
according to data from the three major photometric/spectroscopic surveys 
(ECAS (Eight-Color Asteroid Analysis, \citet{Tholen_1989}), 
SMASS (Small Main Belt Spectroscopic Survey, 
\citet{Bus_2002a,Bus_2002b}), and 
S3OS2 (Small Solar System Objects Spectroscopic Survey, 
\citet{Lazzaro_2004}): 195 Eurykleia, 238 Hypatia, 910 Annelise, and 1189 
Terentia.  The WISE survey \citep{Masiero_2012} estimates that these
objects have diameters of 85.7, 148.5. 47.1, and 55.9 km, respectively.
With the possible and very doubtful exception of Annelise and Terentia,
the other objects are too big to be realistically assumed to be 
members of a Ceres family.  Apart from a small family of 79 members 
around Terentia \citep{Nesvorny_2015}, 
no major family has been identified around the largest objects, that appear to
be isolated.  To increase statistics, we turn our attention to 
photometric data from the Sloan Digital Sky 
Survey-Moving Object Catalog data, fourth release (SDSS-MOC4 hereafter, 
\citet{Ivezic_2001}).  Using the classification method of \citet{DeMeo_2013}
in the $gri$ slope and $z' -i'$ colors domain, we identified 23 C-type
photometric candidates in the pristine region.

\begin{figure*}

  \centering
  \begin{minipage}[c]{0.45\textwidth}
    \centering \includegraphics[width=3.in]{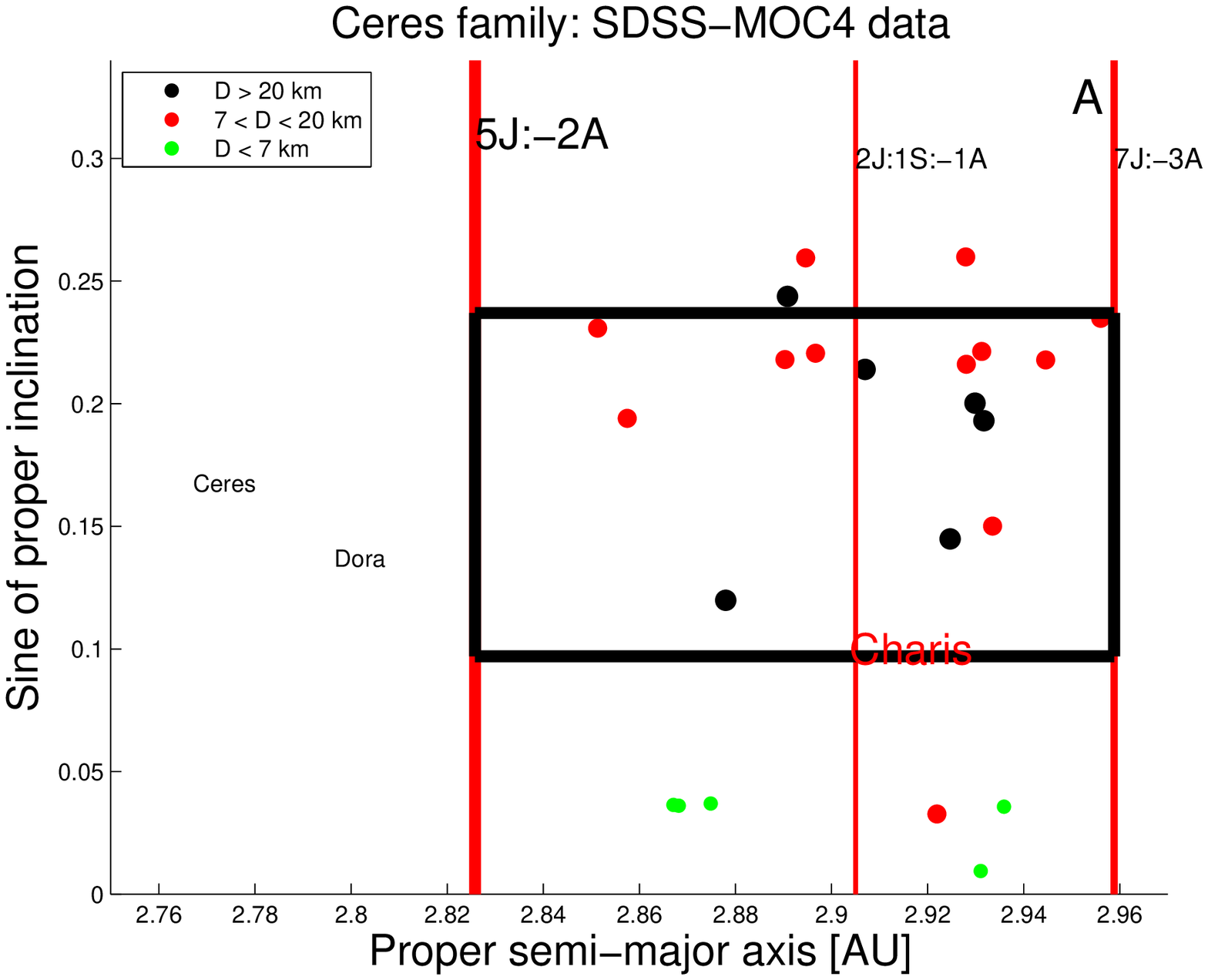}
  \end{minipage}%
  \begin{minipage}[c]{0.45\textwidth}
    \centering \includegraphics[width=3.in]{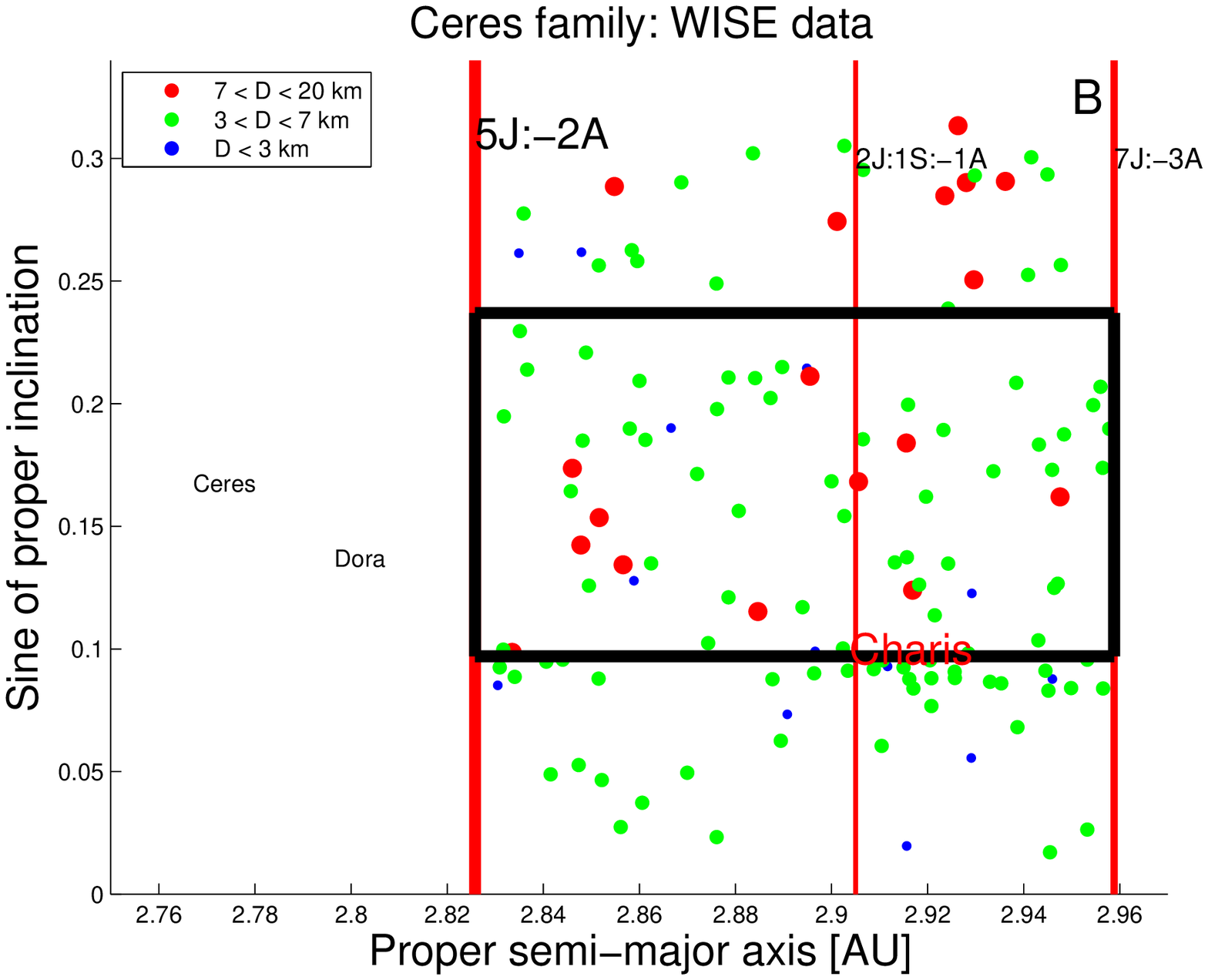}
  \end{minipage}

\caption{An $(a,\sin{(i)})$ projection of 23 C-type SDSS-MOC4 
photometric candidates (panel A) and 130 objects with WISE albedo in the
range from 0.08 to 0.10 (panel B) in the pristine region.  Candidates of 
different diameters are shown with the color code reported in the 
figure legend.  The orbital location of 627 Charis is identified by its
name, in red letters.}
\label{fig: Ceres_sdss_wise}
\end{figure*}

Fig.~\ref{fig: Ceres_sdss_wise}, panel A shows their orbital location in 
the $(a,\sin{(i)})$ plane.  Vertical red lines display the orbital location
of the main mean-motion resonances in the region, including the 2J:1S:-1A
three-body resonance that divides the pristine area in two parts.  
14 of the 23 SDSS C-type asteroids, 60.9\% of the total, can be found in the 
black region where we would expect Ceres family members
to be located.  By contrast, only 6 objects can be
found at lower inclinations, and only 3 at higher values.   Overall,
C-type photometric candidates in the pristine region tend to cluster
at inclination values close to that of Ceres.

To check the statistical significance of this result, we checked
what would be the probability that these 14 asteroids were produced
by a fluctuation of a Poisson distribution (it has been suggested
that the orbital distribution of background asteroids follows
a Poisson distribution \citet{Carruba_2011}).  Following the approach
of \citet{Carruba_2011}, the probability that the expected number of 
$k = 14$ occurrences in a given interval is produced by the Poissonian
statistics is given by 

\begin{equation}
f(k,\lambda)= \frac{{\lambda}^{k} e^{-\lambda}}{k!},
\label{eq: poisson}
\end{equation}

\noindent where $\lambda$ is a positive real number, equal to the
expected number of occurrences in the given interval.  Here we used
for $\lambda$ the number of SDSS candidates in the black region 
($N_{SDSS} =14$), weighted to account for the volume occupied by the 
region ($V_{Reg}$) with respect to the total volume considered in our 
analysis ($V_{Tot}$, we considered asteroids with $\sin{(i)} < 0.35$).  
This is given by:

\begin{equation}
\lambda = N_{SDSS} \frac{V_{Reg}}{V_{Tot}},
\label{eq: lambda}
\end{equation}

\noindent For our considered black region, this yield $\lambda = 5.6$,
that corresponds to a probability to observe 14 bodies of just 0.13\%,
quite less than the null hypothesis level of 1.0\% of the data being drawn
from the Poisson distribution.  Similar analysis performed for uniform
and Gaussian distributions (see \citet{Carruba_2011} for details on the
procedures) also failed to fulfill the null hypothesis, which suggests
that the observed concentration of asteroids at inclinations close to those
of Ceres should be statistically significant.

To increase the number of possible Ceres members, we also look for 
objects in the WISE database whose geometric albedo $p_V$ is between 
0.08 and 0.10, the range of albedo
values observed for more than 90\% of the surface material at Ceres
by the Dawn spacecraft \citep{Li_2015}.  While C-type objects with these
values of WISE albedo could be associated with a Ceres family, other
spectral types such as X, D, L, K, and even some S also have albedo
values in this range, so this data must be considered with some caution.
Nevertheless, it provides useful preliminary information.  After eliminating
objects belonging to spectral types other than C (13 bodies), members of 
the Terpsichore and Terentia families (7 asteroids), and of the 
X-type group Charis (5 objects), we were left with a sample of 
130 asteroids in that range of albedos.  The fact that only 4 C-type 
candidates were members of the Terpsichore family, and 3 of Terentia, 
3.2\% of the total, seems to indicate that local sources may not explain 
all the local population of C-type candidates.
Since the other C-type objects in the region have diameters less than that
of Terpsichore, even if they produced families in the past (of which
there is no trace today), we would expect that to account at most for 10\%
of the observed population of C-type candidates.

Fig.~\ref{fig: Ceres_sdss_wise}, panel B, displays the $(a,\sin{(i)})$ 
distribution of these objects.  Again, 67 asteroids, 51.5\% of the total,
are found in the black region, while 39, (30.0\%) are found at 
lower inclinations and 24 (18.5\%) at higher $i$.   22 of the objects at low
inclinations cluster near the X-type family Charis, and could quite possibly
be associated with its halo. If we eliminate these objects, then
the percentage of albedo candidates in the black region increase to 56.0\%,
a result similar to that from the SDSS-MOC4 data.  To reduce possible 
contaminations from the local large families of Charis, Eos and Koronis, 
which could have small populations of objects with WISE albedos in the 
studied range, we eliminated objects i) with $a$ beyond the 2J:1S:-1A 
mean-motion resonance that could potentially be associated with the Charis and 
Eos halo \citep{Broz_Morby_2013}, and are less likely to have been produced
by collisions on Ceres with low values of $\beta$, and ii) those
objects between the 5J:-2A and the 2J:1S:-1A mean-motion resonances with 
proper eccentricities lower than 0.08 and sine of proper inclinations lower 
than 0.06, that could potentially be members of the Koronis halo 
\citep{Carruba_2016b}.

\begin{figure*}

  \centering
  \begin{minipage}[c]{0.45\textwidth}
    \centering \includegraphics[width=3.in]{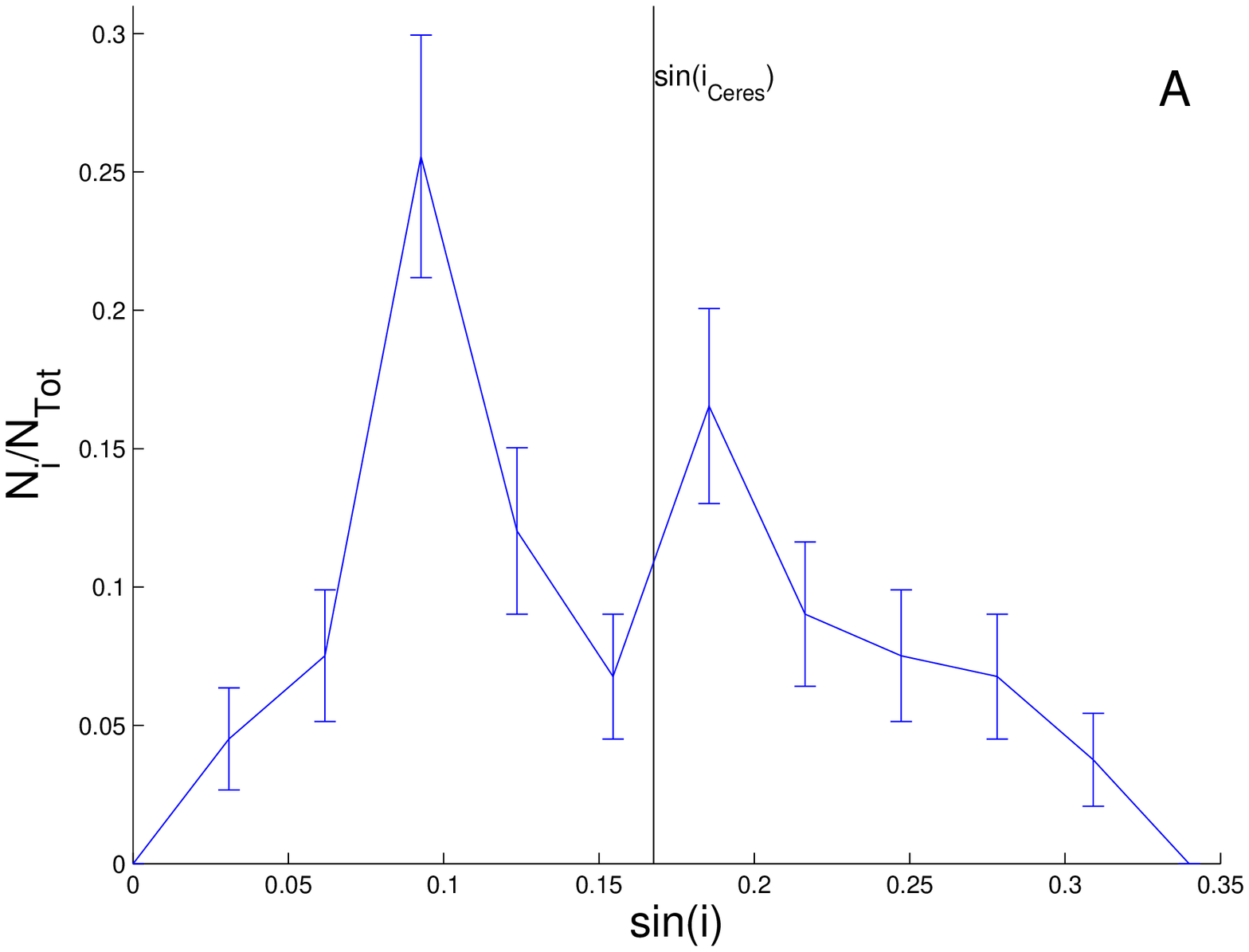}
  \end{minipage}%
  \begin{minipage}[c]{0.45\textwidth}
    \centering \includegraphics[width=3.in]{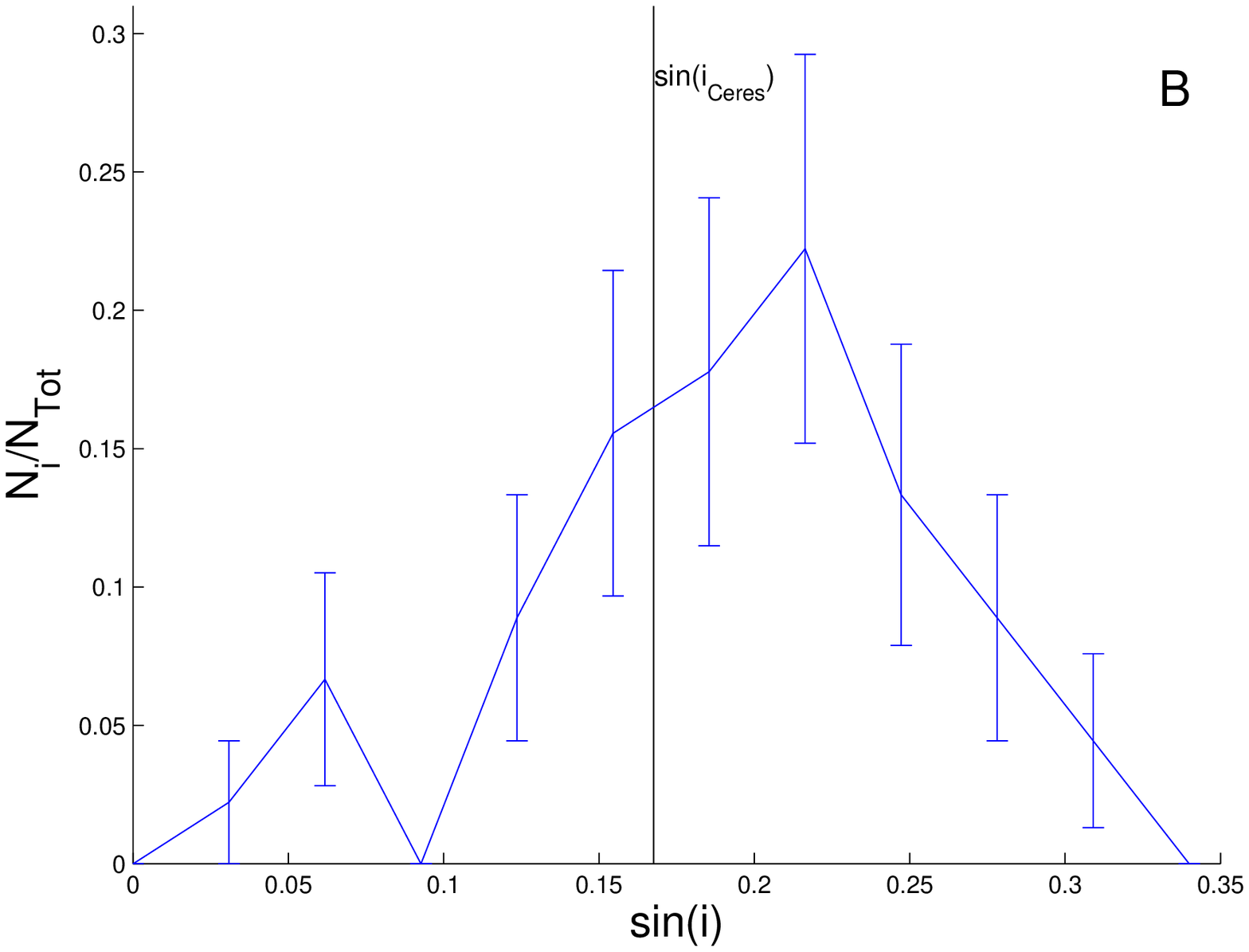}
  \end{minipage}

\caption{A histogram of the distribution of $\sin{(i)}$ values for 133 
and 45 C-type (panel A and B, respectively) albedo candidates in the 
pristine region.  The vertical line marks the value of 
Ceres inclination. Errors are assumed to be proportional to the square 
root of the number of asteroids in each size bin.} 
\label{fig: ceres_histpv}
\end{figure*}

Fig.~\ref{fig: ceres_histpv}  shows a histogram of all the 133 C-type 
albedo candidates in the pristine region (panel A), and of 
the 45 asteroids that satisfied our criteria (panel B).  In the first
panel one can notice a peak at $\sin(i)  \simeq 0.10$ associated with the
local Charis X-type family, that is not visible once objects in the Charis
halo are removed (panel B).  Remarkably, there is a 
peak in the $\sin{(i)}$ distribution that nearly coincides with the 
current value of the inclination of Ceres.  To check for its statistical
significance, and in particular if it could be the product of fluctuations
in a Poisson probability distribution, we used the approach described in 
\citet{Carruba_2011}, Sect. 4.1.  For the $N_{i}$ values of the number of 
asteroids per size bin in the histogram shown in Fig.~\ref{fig: ceres_histpv}, 
panel B, we best-fitted the value of $\lambda$ in Eq.~\ref{eq: poisson}
using the $MATLAB$ routine (where $MATLAB$ stands for Matrix Laboratory)
{\it poissfit}.  We then generated 1000 random populations of asteroids
following the Poisson distribution with the routine {\it poissrnd}, and then
performed a Pearson ${\chi}^2$ test by computing the ${\chi}^2$-like
variable defined as:

\begin{equation}
{\chi}^2 = \sum_{i=1}^{n_{int}} \frac{(q_i -p_i)^2}{q_i},
\label{eq: chi}
\end{equation} 

\noindent where $n_{int}$ is the number of intervals used in the histogram
in Fig.~\ref{fig: ceres_histpv}, panel B with a number of objects larger 
than 0 (9), $q_i$ is the number of real objects in the $i-th$ interval,
and $p_i$ is the number of simulated asteroids in the same $\sin{(i)}$
range.  Assuming that the ${\chi}^2$-like variable follows an incomplete
gamma function probability distribution, the probability of the two
distributions being compatible can be computed considering the minimum
value of ${\chi}^2$ (17.46, in our case) and the number of degrees of
freedom (9, \citet{Press_2001}).  This yields a probability of 0.8\%
of the two distributions being compatible, below the null hypothesis,
which implies that the observed peak should be statistically significant.
Similar results were obtained for the uniform and Gaussian distributions.

While not a conclusive proof of the possible existence of a Ceres family, 
this interesting result represents,
in our opinion, a circumstantial evidence in favor of this 
possibility.  A list of the photometric and albedo Ceres members 
candidates with diameters less than 20 km in the pristine region 
identified in this work is provided in Table~\ref{table: Ceres_members}.

\section{Conclusions}
\label{sec: conc}

Our results could be summarized as follows:

\begin{itemize}

\item We studied the effect that the local dynamics may have had on a possible
Ceres family in the central main belt.  Resonances with Ceres would have 
depleted the population of members at higher inclinations than Ceres itself, 
and close encounters with the dwarf planet would also caused further 
spread of the family, making the identification
of a Ceres dynamical group impossible after timescales of 200-400 My.  
Members of the Ceres family at low inclinations 
would not be easily distinguishable from members of other local large 
Ch families such as Dora and Chloris.

\item We investigated the original spread that a Ceres family may have had 
in the $(a,e,\sin{(i)})$ proper element domain.  Assuming that values
of the initial ejection velocity parameter $\beta =V_{EJ} /V_{esc}$, would 
be in the range observed for other families, we expect that a potentially 
observable population of $D> 3$~km fragments, likely the results of the 
formation of craters on Ceres larger than 300-400 km,
should have been injected into the pristine region of the 
main belt.  Collisional models predict that about 10 such craters could have
formed on Ceres \citep{Marchi_2016}.  While the largest well-defined
basins observed by Dawn on Ceres surface are smaller than 300 km, there is 
topographical evidence that larger impact structures formed in the 
ancient past \citep{Marchi_2016}. The pristine region is 
characterized by a lower number density of 
asteroids, when compared to the central main belt. We argue that Ceres 
family members would be easier to recognize in this region, rather than 
near Ceres itself.

\item We analyzed which objects in the pristine region have taxonomical,
photometric, or albedo data comparable to that of Ceres.  Remarkably, C-type
candidates in this region cluster at inclinations that would be compatible
with the range reached by possible Ceres' fragments.  Among albedo
C-type candidates, we observe a statistically significant peak in the 
$\sin{(i)}$ distribution at values close to that of Ceres itself.  The number
of C-type candidates at lower and higher inclinations is significantly lower,
and no significant local sources can produce the observed distribution 
of C-type candidates.

\end{itemize}

While we believe that we have provided a circumstantial evidence
in favor of a possible Ceres family in this work, we are aware that we 
have not proven its existence beyond any reasonable doubt.  Since Ceres
surface is characterized by a distinct $3\mu m$ absorption band 
\citep{DeSanctis_2015}, observing this feature in some of the proposed 
Ceres family candidates could be a conclusive proof that such family exist, 
or existed in the past.  Should the existence of a Ceres family
be finally proved, future lines of research could investigate the 
possibility of some main belt comets \citep{Hsieh_2015}  being escaped
members of this long-lost asteroid group.

\section*{Acknowledgments}
We are grateful to the reviewer of this paper, Dr. Bojan Novakovi\'{c}, 
for comments and suggestions that improved the quality of this work.
We would like to thank the S\~{a}o Paulo State Science Foundation 
(FAPESP) that supported this work via the grant 14/06762-2, and the
Brazilian National Research Council (CNPq, grant 305453/2011-4).
DN and SM acknowledge support from the NASA SSW and SSERVI programs, 
respectively.  The first author was a visiting scientist at the Southwest 
Research Institute in Boulder, CO, USA, when this article was written.
This publication makes use of data products from the Wide-field 
Infrared Survey Explorer (WISE) and of NEOWISE, which are a joint project of 
the University of California, Los Angeles, and the Jet Propulsion 
Laboratory/California Institute of Technology, funded by the National 
Aeronautics and Space Administration.

\onecolumn
\begin{center}
\begin{longtable}{|c|c|c|c|c|c|c|}
\caption{Photometric and albedo Ceres family candidates in the 
pristine region with $D < 20$~km.  We report 
the asteroid identification, its proper $a,e,\sin{(i)}$, 
its absolute magnitude $H$, the estimated diameter $D$, and if it is 
a photometric (SDSS) or an albedo (WISE) C-type candidate.  
Asteroids in the expected inclination range of the Ceres family are 
marked with a dagger~$(\dagger)$.  The 45 asteroids that satisfied
our selection criteria for Fig.~\ref{fig: ceres_histpv}, panel B, 
have their identifications marked in bold.}
\label{table: Ceres_members}\\
\hline 
Identification & Semi-major axis [au] & Eccentricity & Sine of 
inclination & Absolute magnitude & Diameter [km] & Type \\
\hline
\endfirsthead
\multicolumn{7}{|c|}%
{\tablename\ \thetable\ -- \textit{Continued from previous page}} \\
\hline
Identification & Semi-major axis [au] & Eccentricity & Sine of 
inclination & Absolute magnitude & Diameter [km] & Type \\
\hline
\endhead

\hline \multicolumn{7}{|r|}{{Continued on next page}} \\ \hline
\endfoot

\hline 
\endlastfoot

2082             & 2.921907 & 0.193664 & 0.032701 & 13.30 & 17.12 & SDSS \\
5994~$(\dagger)$ & 2.851295 & 0.184111 & 0.230810 & 11.98 & 17.80 & SDSS \\
6671             & 2.894602 & 0.135151 & 0.259412 & 12.54 & 13.75 & SDSS \\
6721             & 2.928060 & 0.160947 & 0.290183 & 12.40 & 14.93 & WISE \\
8560~$(\dagger)$ & 2.956013 & 0.055937 & 0.234792 & 12.30 & 18.95 & SDSS \\
{\bf 8706}~$(\dagger)$ & 2.847802 & 0.081223 & 0.142328 & 13.30 & 9.59 & WISE \\
11771~$(\dagger)$& 2.916868 & 0.182122 & 0.123918 & 13.90 &  7.60 & WISE \\
12356            & 2.935904 & 0.069843 & 0.035674 & 13.10 &  6.01 & SDSS \\
12986~$(\dagger)$& 2.930536 & 0.071421 & 0.192957 & 13.90 &  7.77 & WISE \\
{\bf 14037}~$(\dagger)$& 2.846048 & 0.113655 & 0.173687 & 12.80 &12.86 & WISE \\
14338~$(\dagger)$& 2.944557 & 0.079173 & 0.217839 & 12.40 &  8.99 & SDSS \\
14531~$(\dagger)$& 2.833541 & 0.136888 & 0.098603 & 13.80 &8.12 & WISE \\
16885~$(\dagger)$& 2.890294 & 0.089698 & 0.218012 & 12.90 & 15.54 & SDSS \\
18198~$(\dagger)$& 2.928039 & 0.165372 & 0.216076 & 13.60 &  9.67 & SDSS \\
{\bf 18463}      & 2.856107 & 0.121585 & 0.027381 & 14.40 &  5.89 & WISE \\
{\bf 20094}~$(\dagger)$& 2.861264 & 0.126932 & 0.185285 & 14.40 & 6.16 & WISE \\
20095            & 2.867089 & 0.044780 & 0.036435 & 14.44 &  5.73 & SDSS \\
20965~$(\dagger)$& 2.929609 & 0.106764 & 0.250502 & 13.20 & 10.62 & WISE \\
21714~$(\dagger)$& 2.896649 & 0.056614 & 0.220559 & 12.50 &  8.27 & SDSS \\
22540            & 2.874836 & 0.064495 & 0.036984 & 14.00 &  3.60 & SDSS \\
23000~$(\dagger)$& 2.931225 & 0.042298 & 0.221314 & 13.90 & 14.06 & SDSS \\
23136            & 2.927916 & 0.115123 & 0.259867 & 13.50 &  8.05 & SDSS \\
23432            & 2.910407 & 0.099806 & 0.060429 & 15.00 &  4.32 & WISE \\
26085            & 2.847349 & 0.071938 & 0.052672 & 14.20 &  6.12 & WISE \\
28297~$(\dagger)$& 2.933505 & 0.149849 & 0.150057 & 13.60 &  9.36 & SDSS \\
29514            & 2.936161 & 0.193732 & 0.290663 & 13.20 & 10.40 & WISE \\
{\bf 29906}~$(\dagger)$& 2.884096 & 0.043687 & 0.210437 & 14.00 & 6.84 & WISE \\
31232~$(\dagger)$& 2.857460 & 0.055909 & 0.194055 & 14.40 &  7.27 & SDSS \\
32421~$(\dagger)$& 2.915570 & 0.131466 & 0.183926 & 13.70 &  8.45 & WISE \\
{\bf 32586}~$(\dagger)$& 2.905601 & 0.142687 & 0.168171 & 12.80 &11.24 & WISE \\
33202            & 2.931046 & 0.058748 & 0.009448 & 14.46 &  5.68 & SDSS \\
34129            & 2.868182 & 0.042556 & 0.036113 & 15.00 &  4.11 & SDSS \\
35533~$(\dagger)$& 2.902357 & 0.098794 & 0.100204 & 14.50 &5.91 & WISE \\
38466            & 2.860576 & 0.046388 & 0.037292 & 15.10 &  4.09 & WISE \\
42544            & 2.914965 & 0.087749 & 0.092480 & 14.70 &  4.85 & WISE \\
42710~$(\dagger)$& 2.939916 & 0.070241 & 0.191939 & 13.90 &  7.49 & WISE \\
{\bf 43235}~$(\dagger)$& 2.835135 & 0.069649 & 0.229597 & 14.00 & 6.72 & WISE \\
44357            & 2.938663 & 0.176620 & 0.068078 & 14.50 &  5.65 & WISE \\ 
46645            & 2.944500 & 0.159099 & 0.091087 & 14.00 &  6.97 & WISE \\
{\bf 47651}~$(\dagger)$& 2.836642 & 0.070337 & 0.213856 & 14.60 & 5.36 & WISE \\
{\bf 47862}~$(\dagger)$& 2.887313 & 0.087727 & 0.202313 & 14.30 & 6.29 & WISE \\
{\bf 51050}~$(\dagger)$& 2.876180 & 0.169556 & 0.197835 & 14.00 & 6.87 & WISE \\
51338~$(\dagger)$& 2.947534 & 0.178943 & 0.161967 & 13.10 & 11.14 & WISE \\
51707~$(\dagger)$& 2.893943 & 0.194367 & 0.117033 & 14.50 &5.82 & WISE \\
{\bf 53875}~$(\dagger)$& 2.895506 & 0.046369 & 0.211182 & 13.70 & 8.32 & WISE \\
54065~$(\dagger)$& 2.948372 & 0.120630 & 0.187491 & 15.00 &  4.25 & WISE \\
54930~$(\dagger)$& 2.915907 & 0.102511 & 0.199583 & 14.90 &  4.42 & WISE \\
54966~$(\dagger)$& 2.928465 & 0.101667 & 0.098231 & 14.50 &5.30 & WISE \\
55903            & 2.941553 & 0.124014 & 0.300482 & 14.50 &  5.68 & WISE \\
57676            & 2.840613 & 0.081003 & 0.094847 & 15.40 &  3.51 & WISE \\
57695            & 2.953239 & 0.067685 & 0.026310 & 15.20 &  4.28 & WISE \\
{\bf 61674}~$(\dagger)$& 2.856602 & 0.145630 & 0.134328 & 13.70 &8.21 & WISE \\
65774            & 2.876079 & 0.084274 & 0.023299 & 15.40 &  3.89 & WISE \\
66046            & 2.923553 & 0.054957 & 0.284802 & 13.80 &  7.83 & WISE \\
{\bf 66648}~$(\dagger)$& 2.857988 & 0.126132 & 0.189854 & 15.50 & 3.57 & WISE \\
67333~$(\dagger)$& 2.884682 & 0.186828 & 0.115222 & 14.00 &7.41 & WISE \\
{\bf 71263}~$(\dagger)$& 2.851656 & 0.166470 & 0.153571 & 13.90 & 7.29 & WISE\\
72691~$(\dagger)$& 2.954464 & 0.092571 & 0.199377 & 14.40 &  5.88 & WISE \\
74398            & 2.926309 & 0.113313 & 0.313344 & 13.20 &  9.78 & WISE \\
{\bf 76814}      & 2.854806 & 0.079048 & 0.288523 & 13.90 &  7.56 & WISE \\
77346            & 2.925627 & 0.114702 & 0.090742 & 15.30 &  3.93 & WISE \\
77873~$(\dagger)$& 2.943052 & 0.183550 & 0.103473 & 14.10 &6.63 & WISE \\
{\bf 82715}~$(\dagger)$& 2.831806 & 0.085409 & 0.194841 & 13.90 & 6.99 & WISE \\
{\bf 82897}~$(\dagger)$& 2.878531 & 0.056788 & 0.210677 & 14.50 & 5.69 & WISE \\
83337~$(\dagger)$& 2.831666 & 0.098578 & 0.099738 & 14.30 &6.24 & WISE \\
83707            & 2.830936 & 0.085110 & 0.092492 & 14.00 &  6.80 & WISE \\
{\bf 85114}      & 2.835897 & 0.132118 & 0.277561 & 15.10 &  4.20 & WISE \\
88573~$(\dagger)$& 2.919644 & 0.132783 & 0.162074 & 14.40 &  5.92 & WISE \\
88723            & 2.834080 & 0.088113 & 0.088653 & 14.90 &  4.69 & WISE \\
89727            & 2.945443 & 0.101456 & 0.017127 & 15.40 &  3.75 & WISE \\
{\bf 91870}~$(\dagger)$& 2.845707 & 0.116806 & 0.164315 & 14.70 &5.11 & WISE \\
91997~$(\dagger)$& 2.921480 & 0.078832 & 0.113695 & 14.80 &5.09 & WISE \\
{\bf 97020}       & 2.901140 & 0.058409 & 0.274274 & 14.10 &  7.04 & WISE \\
103529            & 2.920771 & 0.052655 & 0.088095 & 15.50 &  3.36 & WISE \\
103923~$(\dagger)$&2.933675 & 0.076817 & 0.172426 & 15.40 &  3.61 & WISE \\
104028~$(\dagger)$&2.923278 & 0.054426 & 0.189286 & 14.60 &  5.39 & WISE \\
{\bf 105129}~$(\dagger)$&2.862443 & 0.199475 & 0.134897 & 15.00 & 4.70 & WISE \\
108386~$(\dagger)$&2.956425 & 0.080106 & 0.173829 & 15.20 &  4.10 & WISE \\
111331            &2.917046 & 0.091109 & 0.083911 & 15.20 &  4.21 & WISE \\
111394~$(\dagger)$&2.946302 & 0.090717 & 0.124884 & 14.50 &  5.73 & WISE \\
112976~$(\dagger)$&2.878561 & 0.086200 & 0.121019 & 14.90 &4.56 & WISE \\
113564            &2.851523 & 0.085304 & 0.087892 & 14.80 &  4.78 & WISE \\
119722            &2.852195 & 0.072637 & 0.046523 & 15.30 &  3.92 & WISE \\
120043~$(\dagger)$&2.927114 & 0.071038 & 0.193386 & 14.60 &  5.40 & WISE \\
121281            &2.844024 & 0.150695 & 0.095594 & 15.00 &  4.65 & WISE \\
125857            &2.896361 & 0.102890 & 0.090084 & 15.70 &  3.28 & WISE \\
{\bf 128714}      &2.868729 & 0.041501 & 0.290199 & 14.40 &  5.81 & WISE \\
{\bf 129048}~$(\dagger)$&2.848874 & 0.132366 & 0.220818 & 14.90 & 4.71 & WISE\\
{\bf 130023}~$(\dagger)$&2.880666 & 0.139291 & 0.156271 & 14.80 & 5.14 & WISE \\
{\bf 131267}~$(\dagger)$&2.848184 & 0.130562 & 0.184916 & 14.90 & 4.57 & WISE \\
133841            &2.932969 & 0.065715 & 0.086641 & 14.90 &  4.47 & WISE \\
139184~$(\dagger)$&2.915685 & 0.191200 & 0.137376 & 15.60 &  3.20 & WISE \\
139793            &2.953216 & 0.076844 & 0.095644 & 15.30 &  3.72 & WISE \\
140007            &2.841540 & 0.052047 & 0.048892 & 14.60 &  5.45 & WISE \\
{\bf 140769}~$(\dagger)$&2.860013 & 0.068603 & 0.209332 & 15.20 & 4.21 &WISE \\
{\bf 143166}      &2.869961 & 0.101913 & 0.049494 & 15.30 &  3.81 & WISE \\
144712~$(\dagger)$&2.915620 & 0.171867 & 0.019665 & 15.90 &  2.85 & WISE \\
{\bf 144715}~$(\dagger)$&2.872016 & 0.110133 & 0.171309 & 15.40 & 3.73 &WISE \\
{\bf 145325}~$(\dagger)$&2.849511 & 0.122737 & 0.125786 & 15.20 & 4.14 &WISE \\
{\bf 145392}      &2.902675 & 0.059978 & 0.305121 & 14.80 &  5.07 & WISE \\
155547            &2.935295 & 0.076924 & 0.086005 & 15.30 &  3.88 & WISE \\
{\bf 159263}      &2.851540 & 0.121195 & 0.256403 & 15.40 &  3.66 & WISE \\
166760            &2.945129 & 0.099172 & 0.082949 & 15.00 &  4.22 & WISE \\
{\bf 169161}      &2.889445 & 0.116778 & 0.062533 & 15.90 &  3.04 & WISE \\
{\bf 172377}~$(\dagger)$&2.902666 & 0.179680 & 0.154193 & 14.80 & 4.62 &WISE \\
174489            &2.929807 & 0.132278 & 0.293032 & 14.90 &  4.48 & WISE \\
176347~$(\dagger)$&2.918230 & 0.086535 & 0.126204 & 14.80 &  4.77 & WISE \\
178669            &2.956492 & 0.110930 & 0.083905 & 15.50 &  3.57 & WISE \\
181605~$(\dagger)$&2.87433& 0 0.158614&0.102414 & 15.30 &  4.01 & WISE \\
182687            &2.910644 & 0.119178 & 0.095378 & 15.50 &  3.38 & WISE \\
184356            &2.908767 & 0.107959 & 0.091759 & 15.50 &  3.43 & WISE \\
184487            &2.945952 & 0.095575 & 0.087735 & 16.10 &  2.61 & WISE \\
189495            &2.944911 & 0.133490 & 0.293474 & 14.30 &  6.32 & WISE \\
191486            &2.947722 & 0.145902 & 0.256494 & 14.60 &  5.36 & WISE \\
192044            &2.940902 & 0.074702 & 0.252544 & 14.70 &  5.04 & WISE \\
197144            &2.925675 & 0.105006 & 0.088102 & 15.00 &  4.34 & WISE \\
198403~$(\dagger)$&2.906537 & 0.088144 & 0.185480 & 15.50 &  3.37 & WISE \\
198808~$(\dagger)$&2.924283 & 0.073838 & 0.238708 & 14.60 &  5.27 & WISE \\
{\bf 201824}      &2.876024 & 0.082627 & 0.248968 & 14.90 &  4.54 & WISE \\
214826            &2.949845 & 0.113408 & 0.084030 & 15.80 &  3.03 & WISE \\
218273            &2.887695 & 0.078069 & 0.087647 & 15.50 &  3.70 & WISE \\
218776~$(\dagger)$&2.924260 & 0.057623 & 0.134816 & 15.60 &  3.38 & WISE \\
{\bf 222080}~$(\dagger)$&2.866610 & 0.149922 & 0.190036 & 16.10 & 2.68 & WISE\\
227004~$(\dagger)$&2.913158 & 0.058757 & 0.135286 & 15.20 &  3.93 & WISE \\
{\bf 231791}~$(\dagger)$&2.889728 & 0.157418 & 0.214961 & 14.90 & 4.84 & WISE\\
232724            &2.903389 & 0.120023 & 0.091075 & 15.50 &  3.66 & WISE \\
235975            &2.920486 & 0.101497 & 0.095416 & 15.30 &  3.77 & WISE \\
236004~$(\dagger)$&2.943159 & 0.076640 & 0.183347 & 15.30 &  3.67 & WISE \\
238734~$(\dagger)$&2.957799 & 0.056591 & 0.189721 & 15.60 &  3.26 & WISE \\
{\bf 239518}      &2.859584 & 0.120313 & 0.258105 & 15.40 &  3.50 & WISE \\
239993~$(\dagger)$&2.955953 & 0.053980 & 0.206900 & 15.20 &  3.98 & WISE \\
{\bf 240384}      &2.834916 & 0.123127 & 0.261325 & 16.30 &  2.45 & WISE \\
243283            &2.890796 & 0.122268 & 0.073358 & 16.40 &  2.45 & WISE \\
243302~$(\dagger)$&2.938414 & 0.125304 & 0.208466 & 15.90 &  3.06 & WISE \\
243774            &2.906543 & 0.130365 & 0.295275 & 14.90 &  4.43 & WISE \\
243854~$(\dagger)$&2.947074 & 0.198642 & 0.126632 & 15.70 &  3.30 & WISE \\
{\bf 244955}~$(\dagger)$&2.899970 & 0.057671 & 0.168345 & 15.50 & 3.50 & WISE\\
{\bf 245480}      &2.858417 & 0.093258 & 0.262532 & 15.60 &  3.50 & WISE \\
246409            &2.929076 & 0.145512 & 0.055515 & 15.90 &  2.84 & WISE \\
246466            &2.830511 & 0.084131 & 0.085150 & 16.10 &  2.64 & WISE \\
246720            &2.911652 & 0.076910 & 0.092850 & 16.10 &  2.82 & WISE \\
{\bf 247195}      &2.847924 & 0.120740 & 0.261754 & 15.90 &  2.98 & WISE \\
248035~$(\dagger)$&2.929195 & 0.139209 & 0.122659 & 15.80 &  3.00 & WISE \\
248581            &2.920784 & 0.100786 & 0.076717 & 15.70 &  3.20 & WISE \\
249201            &2.916190 & 0.077005 & 0.087743 & 15.70 &  3.08 & WISE \\
{\bf 249283}~$(\dagger)$&2.894856 & 0.105884 & 0.214428 & 16.10 &2.72 & WISE\\
250010~$(\dagger)$&2.896574 & 0.126868 & 0.098989 & 16.20 &  2.58 & WISE \\
260891~$(\dagger)$&2.945898 & 0.111076 & 0.172986 & 15.80 &  3.22 & WISE \\
{\bf 261489}      &2.883625 & 0.146659 & 0.302045 & 15.60 &  3.21 & WISE \\
{\bf 268622}~$(\dagger)$&2.858875 & 0.075933 & 0.127763 & 16.20 &2.58 & WISE\\
\hline
\end{longtable}
\end{center}
\twocolumn

\bsp

\label{lastpage}

\end{document}